\documentclass[twocolumn,showpacs,preprintnumbers,amsmath,amssymb]{revtex4}

\usepackage{graphicx}
\usepackage{dcolumn}
\usepackage{bm}
\usepackage{graphics}
\usepackage{amsmath}
\usepackage{amssymb}
\usepackage{amscd}
\usepackage{afterpage}
\usepackage{float,times}
\usepackage{subfigure}
\usepackage{rotating}
\usepackage{multirow}
\usepackage{fancyheadings}
\usepackage{epsfig}
\usepackage{theorem}
\usepackage{moreverb}
\usepackage{euscript}
\usepackage{psfrag}

\begin{document}

\title{Systematic Study Of Leptonic Mixing In A Class Of $\mathbf{SU_H(2)}$ Models.}

\author{K. L. McDonald}\email{k.mcdonald@physics.unimelb.edu.au}

\author{B. H. J. McKellar}
 \email{b.mckellar@physics.unimelb.edu.au}
\affiliation{%
School of Physics, Research Centre for High Energy Physics, The
University of Melbourne, Victoria, 3010, Australia\\
}%

\date{\today}

\begin{abstract}
We perform a systematic analysis of the PMNS matrices which arise when
one assigns the three generations of leptons to the $2\oplus 1$
representation of a horizontal $SU_H(2)$ symmetry. This idea has been
previously explored by Kuchimanchi and Mohapatra. However, we assume
$(i)$ the neutrino mass matrix results from leptonic couplings to
$SU_L(2)$ triplet scalar fields and $(ii)$ hierarchies exist amongst
lepton mass matrix elements which result from couplings to scalar
fields with different $SU_H(2)$ charges. Of the sixteen candidate PMNS
matrices which result it is found that only one is both predictive and
possesses a leading order structure compatible with experimental
data. The relevant neutrino mass matrix displays the symmetry
$L_e-L_\mu-L_\tau$ to leading order and we explore the perturbations
required to produce a realistic lepton spectrum. The effective mass in
neutrinoless double beta decay is required to lie in the range
$<m>/(10^{-2}\mathrm{eV})\in[0.7,2.5]$, which is just below current
experimental bounds. $U_{e3}$ is non-zero but not uniquely
determined.
\end{abstract}

\maketitle

\section{\label{sec:non_dem_intro}INTRODUCTION}
The search for a possible connection between the observed families of
fermions, the properties possessed by the fermions, and an underlying
horizontal symmetry has a long
history (see for
example~\cite{Yanagida:1979gs,Maehara:1979kf,Yanagida:1980wd,Yanagida:1980xx,Yanagida:1980xy,
Wilczek:1977uh,Wilczek:1978xi,Jones:1981ep,Shaw:1992gk,Davidson:1998vr,Sarkar:1983tm,
Ray:2001tm,Berezhiani:2001mh,Antonelli:2002zf,Kribs:2003jn,Pomarol:1995xc,Barbieri:1995uv,Barbieri:1996ae,Falcone:2001dh,Anoka:2005vb}).
In more recent years, the experimental observation of neutrino mixing
has lead to an enormous amount of research into a link between the
bi-large mixing in the Pontecorvo-Maki-Nakagawa-Sakata (PMNS) matrix
and a horizontal symmetry. In particular, some recent
works~\cite{Kuchimanchi:2002yu,Kuchimanchi:2002fi,Dutta:2003ps} have attempted
to relate the observed values of the PMNS matrix to the structure
enforced upon the lepton mass matrices by an
$SU_H(2)$ symmetry.

In this paper we
perform a systematic analysis of the PMNS mixing matrices that arise when the three lepton families are assigned to the
$2\oplus 1$-representations of $SU_H(2)$, and the neutrino mass
matrix results from couplings of the $SU_L(2)$ doublet leptons to $SU_L(2)$ triplet scalar fields.

The main purpose of the study is to attempt to relate the observed
PMNS matrix structure to a high energy $SU_H(2)$ symmetry. To prevent
the PMNS matrix from becoming too general, we assume that hierarchies exist amongst
contributions to the lepton mass matrices which result from scalars
with different horizontal quantum numbers. These
hierarchies dictate certain structures upon the lepton mass matrices
and thus provide distinct signatures of the horizontal symmetry.

The assumed hierarchies result in sixteen different leading order PMNS mixing matrices. The majority of these are found to be incompatible with the experimentally extracted values of the lepton mixing matrix due to, e.g., texture zeros in unwanted locations. Of the remaining matrices only one is found to have any predictive quality. The identification of this candidate PMNS matrix is the main result of this study.

This mixing matrix results from a neutrino mass matrix possessing the
non-standard lepton number symmetry $L_e-L_\mu-L_\tau$. It predicts
$U_{e3}=0$ and maximal solar mixing to leading order, whilst the
atmospheric mixing angle remains a free parameter. First order
perturbations, which result from sub-dominant lepton mass matrix
entries, prove adequate to produce realistic charged lepton and
neutrino mass spectra. They also perturb the solar mixing angle to its
preferred non-maximal value and induce non-zero corrections to
$U_{e3}$ which are small but not constrained. The experimental data
directly constrains the parameters which enter into the effective mass
in neutrinoless double beta decay. Consequently it is required to lie
in the range,
\begin{eqnarray}
<m>/(10^{-2}\mathrm{eV})\in[0.7,2.5],
\end{eqnarray}
placing it just below current experimental bounds.

The layout of this paper is as follows. In
Section~\ref{sec:non_dem_overview} we provide some background and
motivations for the present investigation. The scalar sector is
described in Section~\ref{sec:non_dem_scalar_content} whilst
Section~\ref{sec:non_dem_assumptions_in_study} contains the study of
candidate lepton mixing matrices. This section contains our main
results. Some comments on the scale of $SU_H(2)$ symmetry breaking are
offered in Section~\ref{sec:non_dem_scalars_bosons} and we conclude in Section~\ref{sec:non_dem_conclusion}.
%----------------------------------------------------------------%
\section{Background And Motivations\label{sec:non_dem_overview}}
In the present work we shall study the lepton mixing that occurs in
models with an $SU_H(2)$
symmetry, under the assumption that the dominant contribution to the
neutrino mass matrix results from couplings to $SU_L(2)$ triplets. This could occur in models of the type
proposed by Kuchimanchi and Mohapatra
(KM)~\cite{Kuchimanchi:2002yu,Kuchimanchi:2002fi} by extending
the scalar sector to include the necessary $SU_L(2)$ triplet
fields. In this case the horizontal symmetry breaking scale must be
large enough, or the relevant coupling constants small enough, to
render the see-saw contributions to the neutrino mass matrix contributions sub-dominant.

Alternatively one may consider
a model where the neutrino field $N_R$, an $SU_H(2)$ doublet
of right-chiral neutrinos, is not present. The
field $N_R$ was
necessary in the KM models to remove the global anomaly. One could instead
consider a model with a gauge group
$\mathcal{G}_{SM}\otimes\mathcal{G}_{SM}'\otimes SU_H(2)$, where
$\mathcal{G}_{SM}$ is the standard model (SM) gauge group,
$\mathcal{G}_{SM}'$ describes
a mirror sector and the horizontal symmetry couples the two
sectors. This would allow the groups $\mathcal{G}_{SM}$ and
$\mathcal{G}_{SM}'$ to each possess an odd number of $SU_H(2)$
doublets whilst the entire theory is free of the global anomaly. In
this way mathematical consistency of the model would not necessitate
$N_R$. We note that an $SU_H(2)$ symmetry communicating with a mirror
sector is always free from the global anomaly in much the same way as
an $SU_H(3)$ which communicates with a mirror sector is free from the
$[SU_H(3)]^3$ anomaly~\cite{Berezhiani:2005ek}.

The two model building strategies suggested here to obtain a dominant electroweak
triplet contribution to the neutrino mass matrix admittedly represent
a further
complication of the KM models. We consider, however, the study of the
lepton mixing that arises under the assumption of triplet dominance in
the KM type models to be of interest for two reasons.

Firstly, the existence of a horizontal $SU_H(2)$ symmetry which manifests itself in the neutrino mass matrix through
couplings to $SU_L(2)$ triplets would surely represent a very direct
link between the physics beyond the SM and the lepton mixing
matrix. This is because the horizontal symmetry may dictate certain
structures directly upon the neutrino Majorana mass matrix, providing
experimentally accessible information about the nature of higher
energy physics. If nature employs a see-saw mechanism to obtain the
light observed neutrinos, the link between an operative symmetry and
the observed lepton mixing matrix may not be as direct. The see-saw
mechanism allows the structure of the heavy Majorana mass
matrix to influence the structure of the mass matrix for the light
neutrinos, ie $M_\nu=-M_D M^{-1}_R M_D^T$, where $M_D$ ($M_R$) is the
Dirac (heavy Majorana) mass matrix. Structures enforced upon $M_D$ and
$M_R$ by a symmetry may not be as readily reconstructed by analysis of
experimental data as a structure imparted directly upon $M_\nu$.

The second reason is borne out by the structure of the neutrino mass
matrix observed when one groups the muon and tauon generations
together under the horizontal symmetry. We shall show in section
\ref{sec:non_dem_scalar_content} that the most general neutrino mass matrix in this case takes the
form
\begin{eqnarray}
M_\nu=\left(\begin{array}{ccc} s_1 & d_1 & d_2 \\d_1 &
t_1 & t_2\\d_2 & t_2 & t_3
\end{array}\right),\label{non_dem_example_mass_matrix}
\end{eqnarray}
where the entries labelled $s$, $d$ and $t$ result from couplings to
$SU_L(2)$ triplets which form singlet, fundamental and adjoint
representations of $SU_H(2)$. To understand our interest in this
structure we make a brief digression.

It is known that the observed bi-large mixing may result from a
perturbation upon a mixing matrix with the bi-maximal
form~\cite{Vissani:1997pa,Barger:1998ta,Baltz:1998ey} (see also~\cite{Gomez-Izquierdo:2006kk})
\begin{eqnarray}
U_{BM}=\left(\begin{array}{ccc} \frac{1}{\sqrt{2}} &
    \frac{1}{\sqrt{2}} & 0 \\-\frac{1}{2} &\frac{1}{2} &
    \frac{1}{\sqrt{2}}\\ \frac{1}{2}  & -\frac{1}{2}  & \frac{1}{\sqrt{2}}
\end{array}\right).\label{non_dem_bi_maximal}
\end{eqnarray} 
There are three particularly interesting leading order mass matrices which lead to bi-maximal
mixing:\\
$\bullet$ The normal hierarchy, $m_3\gg m_{2,1}$, occurs when
\begin{eqnarray}
M_\nu\approx\frac{\sqrt{\Delta m^2_A}}{2}\left(\begin{array}{ccc} 0 & 0 & 0 \\0 &
1 & 1\\0 & 1 & 1
\end{array}\right),\label{non_dem_normal_hierarchy}
\end{eqnarray}
where $\Delta m_A^2$ is the atmospheric mass-squared difference. This matrix possesses the flavour symmetry $L_e$ and results in maximal
mixing in the 2-3 sector.\\
$\bullet$ One obtains an inverted hierarchy, $m_2\approx m_1\gg m_3$,
when
\begin{eqnarray}
M_\nu\approx\sqrt{\frac{\Delta m^2_A}{2}}\left(\begin{array}{ccc} 0 & -1 & 1 \\-1 &
0 & 0\\1 & 0 & 0
\end{array}\right).\label{non_dem_inverted_hierarchy}
\end{eqnarray}
This mass matrix produces exactly the bi-maximal form of the mixing
matrix. It conserves the flavour symmetry $L'=L_e-L_\mu-L_\tau$,
though it is not the most general matrix permitted by this
symmetry. It has received much
attention~\cite{non_dem_lprime_symmetry}.\\
$\bullet$ Quasi-degenerate neutrinos, $m_1\approx m_2 \approx m_3$,
are obtained when:
\begin{eqnarray}
M_\nu\approx m\left(\begin{array}{ccc} 1 & 0 & 0 \\0 &
0 & -1\\0 & -1 & 0
\end{array}\right).\label{non_dem_quasi_degenerate}
\end{eqnarray}
Matrices of this type have been considered previously
in~\cite{Binetruy:1996cs,Babu:2002dz,Hirsch:2003dr}. This matrix
possess an $L_\mu-L_\tau$ symmetry, which brought it to the attention
of the authors of~\cite{Bell:2000vh,Choubey:2004hn,Rodejohann:2005ru}.

It is interesting to note a feature which is common to each of the
mass matrices
(\ref{non_dem_normal_hierarchy}), (\ref{non_dem_inverted_hierarchy})
and (\ref{non_dem_quasi_degenerate}).
Each of these mass matrices can be obtained by imposing a family
symmetry which differentiates in some sense between the electron
generation and the mu and tau generations. The matrix
(\ref{non_dem_normal_hierarchy}) arises from a symmetry operative only
on the electron neutrino, the $L'$ symmetry of
(\ref{non_dem_inverted_hierarchy}) differentiates clearly between
$\nu_e$ and $\nu_{\mu,\tau}$, whilst the $L_\mu-L_\tau$ symmetry is
operative only on $\nu_\mu$ and $\nu_\tau$.

It is this common feature of these contending leading order neutrino
mass matrices that motivates our interest in a Majorana mass matrix
with the structure given by (\ref{non_dem_example_mass_matrix}). We
see that if one sets $s_1,d_{1,2}=0$ in
(\ref{non_dem_example_mass_matrix}) the non-zero entries in $M_\nu$
correspond to those of the normal hierarchy
(\ref{non_dem_normal_hierarchy}). In
(\ref{non_dem_example_mass_matrix}) the entries $s_1$ and $d_{1,2}$
arise from couplings to different scalars than the one that produces
the entries $t_{1,2,3}$. The dominant mass matrix structure may be
attributed to the absence, or weaker
coupling to, certain scalar representations of the gauge
group. Similarly by setting $s_1,t_{1,2,3}=0$ one obtains a
mass matrix with the same zeros as
(\ref{non_dem_inverted_hierarchy}) so again this structure may be
obtained by consideration of the scalar content of a given model. The quasi-degenerate spectrum can
also be approximated by (\ref{non_dem_example_mass_matrix}) if
$d_{1,2}=0$ and $t_3>t_{1,2}$, though the correlation between the scalar content
required to produce (\ref{non_dem_example_mass_matrix}) and the
leading order structure of $M_\nu$ is more tenuous in this case.

More details regarding the correlation between the structure of
(\ref{non_dem_example_mass_matrix}) and the scalar content of a given model
appear in section \ref{sec:non_dem_scalar_content}. At this stage we wish merely to make
the point that a matrix with this structure can be related to those
which reproduce the inverted and normal hierarchies in a simple
fashion, and that the motivation for the present work stems
largely from this observation.

For the present study we shall not need to distinguish between the
case where the horizontal symmetry communicates with a mirror sector
or that when one includes the states $N_R$ to remove the global
anomaly. However the constraints on the mass of the $SU_H(2)$ gauge
bosons will differ in each of these cases and we shall comment on this
in Section~\ref{sec:non_dem_scalars_bosons}.
%-------------------------------------------------------------------------------%
\section{Scalar Fields Required For General Lepton Mass Matrices\label{sec:non_dem_scalar_content}}
In this section we obtain the set of scalar fields, forming
representations of $\mathcal{G}=SU_L(2)\otimes U_Y(1)\otimes SU_H(2)$,
required to produce the most general lepton mass matrices. The fermion
content is
\begin{eqnarray}
& &\Psi=(\psi_{\mu'},\psi_{\tau'})\sim(2,-1,2), \nonumber\\
& &\psi_{e'}\sim(2,-1,1),\nonumber\\
& &E_R=(\mu_R',\tau_R')\sim(1,-2,2),\nonumber\\
& &e_R'\sim(1,-2,1),\label{non_dem_fermion_content_scalar_section}
\end{eqnarray}
with the primes indicating weak eigenstates.
We have not included any right-chiral neutrino states, however it should be remembered that one may include
these states. As in the original KM model we assign the quarks to trivial representations of $SU_H(2)$. With the above lepton representations the most
general charged leptons mass Lagrangian is
\begin{eqnarray}
\mathcal{L}_{M_e}&=&g\bar{\Psi}\Phi e_{R}'+
g_1\bar{\Psi}\phi vE^T_{R}+g_2\bar{\psi}_{e'}\phi e_R'\nonumber\\
& &+g_3\bar{\psi}_{e'} \Phi\epsilon E^T_{R} +g_4\bar{\Psi}\chi E^T_{R } +\mathrm{h.c.},
\end{eqnarray}
where $\epsilon$ is the two dimensional anti-symmetric tensor and the scalars form the representations
\begin{eqnarray}
\phi\sim(2,1,1),\mkern15mu\Phi\sim(2,1,2),\mkern15mu\chi\sim(2,1,3).
\end{eqnarray}
We denote the scalar vacuum expectation values (VEV's) as
\begin{eqnarray}
<\chi>=\left(\begin{array}{c} 0\\
    1\end{array}\right)\otimes\left(\begin{array}{cc}
    \frac{1}{\sqrt{2}}<\chi_2>& <\chi_1> \\<\chi_3>&-\frac{1}{\sqrt{2}}<\chi_2>
\end{array}\right),
\end{eqnarray}
\begin{eqnarray}
<\phi>=\left(\begin{array}{c} 0\\
    v\end{array}\right),\mkern20mu
<\Phi>=\left(\begin{array}{cc} 0& 0\\
    \kappa_1&\kappa_2\end{array}\right).
\end{eqnarray}
The mass Lagrangian becomes
\begin{eqnarray}
\mathcal{L}_{M_e}&=&\bar{L}_L'M_{l}L_R'+\mathrm{h.c.} + ....,
\end{eqnarray}
where $L'=(e',\mu',\tau')^T$, the dots represent interaction terms and
\begin{eqnarray}
M_{l}=\left(\begin{array}{ccc} g_2v&-g_3\kappa_2& g_3\kappa_1
    \\g\kappa_1&g_4\frac{<\chi_2>}{\sqrt{2}}+g_1v&g_4<\chi_1>\\g\kappa_2&
    g_4<\chi_3>&-g_4\frac{<\chi_2>}{\sqrt{2}}+g_1v \end{array}\right).
\end{eqnarray}
To simplify notation we denote this by
\begin{eqnarray}
M_{l}=\left(\begin{array}{ccc} S_2&D_3& D_4
    \\D_1&T_2+S&T_1\\D_2&
    T_3&-T_2+S \end{array}\right).\label{non_dem_defining_m_electron_form}
\end{eqnarray}
where the entries labelled $S$, $D$ and $T$ arise from scalars forming
singlet, doublet and triplet representations of $SU_H(2)$.

The most general neutrino mass Lagrangian is
\begin{eqnarray}
\mathcal{L}_{M_\nu}&=&f\bar{\Psi^c}\Delta_{2}\psi_{e'}+f_1\bar{\Psi^c}\Delta_{3}\Psi +\nonumber\\
& &f_2\bar{\psi_{e'}^c}\Delta_{1}\psi_{e'}+\mathrm{h.c.},
\end{eqnarray}
 where
\begin{eqnarray}
\Delta_{1}\sim(3,2,1),\mkern10mu\Delta_{2}\sim(3,2,2),\mkern10mu \Delta_{3}\sim(3,2,3).\label{non_dem_defining_scalar_triplet_gauge_numbers}
\end{eqnarray}
The charge conjugate
fermion fields are defined as $\psi_{e'}^c=\epsilon C\psi_{e'}^*$ and
$\Psi^c=C\epsilon\Psi^*\epsilon$ where $C$ is the Dirac space charge
conjugation matrix. Observe that the
Lagrangian $\mathcal{L}_{M_\nu}$ does not contain a term of the type
$\bar{\Psi}^c\Delta_{1}\Psi$. The scalars must acquire VEV's in the electrically neutral direction
and we may denote these VEV's as \mbox{$<\Delta_1^o>$},
\mbox{$<\Delta_{21}^o>$}, \mbox{$<\Delta_{22}^o>$}, 
\mbox{$<\Delta_{31}^o>$} etc.

The Lagrangian $\mathcal{L}_{M_\nu}$ may be rewritten as
\begin{eqnarray}
\mathcal{L}_{M_\nu}=\bar{L}^c_{\nu'} M_\nu L_{\nu'} + \mathrm{h.c.}+....
\end{eqnarray}
where $L_{\nu'}=(\nu_{e'L},\nu_{\mu' L},\nu_{\tau' L})^T$ and
\begin{eqnarray}
M_{\nu}=\left(\begin{array}{ccc} f_2<\Delta_1^o>&-f<\Delta_{22}^o>& f<\Delta_{21}^o>
    \\-f<\Delta_{22}^o>&-f_1<\Delta_{31}^o> &\frac{f_1}{\sqrt{2}}<\Delta_{32}^o> \\f<\Delta_{21}^o> &
   \frac{f_1}{\sqrt{2}}<\Delta_{32}^o> &f_1<\Delta_{33}^o> \end{array}\right).\nonumber
\end{eqnarray}
Again, to simplify notation we write
\begin{eqnarray}
M_{\nu}=\left(\begin{array}{ccc} s_1&d_1& d_2
    \\d_1&t_1&t_2\\d_2&
    t_2&t_3 \end{array}\right),\label{non_dem_defining_mnu_form}
\end{eqnarray}
with the entries labelled $s$, $d$ and $t$ resulting from couplings to
$SU_L(2)$ triplet fields which form singlet, doublet and triplet
representations of $SU_H(2)$ respectively.

The mass matrices (\ref{non_dem_defining_m_electron_form}) and (\ref{non_dem_defining_mnu_form}) demonstrate the relationship between the structure
of the lepton mass
matrices and the scalar content in the class of models we are
studying. In the next section we study the leading
order PMNS matrices obtained under assumed hierarchies within the
matrices (\ref{non_dem_defining_m_electron_form}) and (\ref{non_dem_defining_mnu_form}).  
%-------------------------------------------------------------------------------%
\section{Determining The PMNS Mixing Matrix\label{sec:non_dem_assumptions_in_study}}
Certain assumptions
regarding the hierarchy of elements within the lepton mass matrices
have been made for the purpose of our analysis. We now discuss these
assumptions.

We study the PMNS
matrices obtained when one assumes hierarchies of the type $|d|>>|s|>>|t|$,
$|T|>>|S|>>|D|$ etc, where we collectively label all entries resulting from
couplings with $SU_H(2)$ singlets, doublets and triplets in the
neutrino (charged lepton) sector by $s$, $d$ and $t$ ($S$, $D$ and $T$)
respectively. Without an assumption of this type the mass matrices
are very general and are unlikely leave a signature of the
horizontal symmetry. The choice of this type of hierarchy rests on the
idea that the given leptons may uniformly couple more strongly/weakly to a given
scalar representation than others, or that there is a hierarchy
amongst the VEV's of the different representations. It
may be unnecessary to include a given scalar in a
realistic model if it is found that mass matrix entries arising from this
particular scalar are not required to produce realistic mixing.
Deviations from this type of hierarchy are considered when it is
either physically motivated or presents no significant complication of
the overall mixing matrix. 

During our analysis we shall retain only two of the three
distinctive scalar representations in each of the lepton sectors. The representations absent from the
analysis may be considered to contribute sub-dominantly to a given mass
matrix or may not be
necessary in model construction. To simplify matters we shall take all mass matrix elements to be real.

The investigations are structured as follows. We begin
by obtaining the leading form of the charged lepton mixing matrix
$U_l$, where
\[U_lM_lM_l^\dagger
U_l^\dagger=\mathrm{diag}(m_e^2,m_\mu^2,m_\tau^2).\]
The matrix $U_l$ is determined for various
assumed hierarchies of the elements of $M_l$. The neutrino mixing
matrix $U_\nu$, where
\[U_\nu M_\nu U_\nu^T=M^d_\nu=\mathrm{diag}(m_1,m_2,m_3),\]
is then determined under the assumed hierarchies of elements of
$M_\nu$. The various combinations of the matrices $U_l$ and $U_\nu$
are then combined to determine the resulting PMNS
matrix $U$, where $U=U_lU_\nu^\dagger$.

We now present an example analysis for the charged lepton and neutrino
sectors. We shall present the details for the case which turns out to
be most interesting. The results for the other cases can be found in
the table below.

The most general charged lepton
mass matrix is
\begin{eqnarray}
M_{l}=\left(\begin{array}{ccc} S_2&D_3& D_4
    \\D_1&T_2+S&T_1\\D_2&
    T_3&-T_2+S \end{array}\right).\label{non_dem_defining_m_electron_form_mu_tau}
\end{eqnarray}
We shall refer to the $SU_L(2)$ doublet scalars $\phi$, $\Phi$ and $\chi$
respectively as
the singlet, doublet and triplet scalars. Consider the case when only
the singlet and triplet scalars are 
retained. In this case the mass matrix is
\begin{eqnarray}
M_{l}=\left(\begin{array}{ccc} S_2&0& 0
    \\0&T_2+S&T_1\\0&
    T_3&-T_2+S \end{array}\right).\label{non_dem_triple_singlet_mass_matrix}
\end{eqnarray}
One cannot take $S\gg T_{1,2,3},S_2$ as this produces $m_\mu\approx
m_{\tau}$. The size of $S_2$ relative to $T_{1,2,3},S$ allows two
possibilities. If one takes $S_2\gg T_{1,2,3},S$ then
\begin{eqnarray}
U_{l}^A=\left(\begin{array}{ccc} 0&c_A&-s_A\\0&
    s_A&c_A \\1&0&0\end{array}\right),\label{non_dem_triple_singlet_Ul_S2_gg}
\end{eqnarray}
whilst for $T_{1,2,3}\gg S,S_2$ or $T_{1,2,3},S\gg S_2$ one has
\begin{eqnarray}
U_{l}^B=\left(\begin{array}{ccc}1&0&0\\ 0&c_B&-s_B\\0&
    s_B&c_B \end{array}\right),\label{non_dem_triple_singlet_Ul_T_gg}
\end{eqnarray}
where we have adopted the notation $c_X=\cos\theta_X$,
$s_X=\sin\theta_X$ and
\begin{eqnarray}
\tan2\theta_A&=&\tan2\theta_B\nonumber\\
&=&\frac{2\{T_3(T_2+S)+T_1(S-T_2)\}}{T_3^2-T^2_1-4T_2S}.
\end{eqnarray}
For the case of $T_{1,2,3}\gg S,S_2$ one may also make the approximation
\begin{eqnarray}
\tan2\theta_B\approx\frac{2T_2}{T_3+T_1}.
\end{eqnarray}

If one instead retains the triplet and doublet scalars the mass matrix is
\begin{eqnarray}
M_{l}=\left(\begin{array}{ccc} 0&D_3& D_4
    \\D_1&T_2&T_1\\D_2&
    T_3&-T_2 \end{array}\right).\label{non_dem_triple_doublet_mass_matrix}
\end{eqnarray}
Under the hierarchy $T\gg D$ one may write
\begin{widetext}
\begin{eqnarray}
M_{l}M_l^\dagger=\left(\begin{array}{ccc} 0&0&0
    \\0&T_1^2+T_2^2&T_2(T_3-T_1)\\0&
    T_2(T_3-T_1)&T_2^2+T_3^2 \end{array}\right)+\left(\begin{array}{ccc} 0&D_1T_2+D_4T_1&D_3T_3-D_4T_2
    \\D_1T_2+D_4T_1&0&0\\D_3T_3-D_4T_2&
    0&0 \end{array}\right) +O(D^2),\label{non_dem_double_triplet_approximate_ml_mldag}
\end{eqnarray}
\end{widetext}
and treat the second matrix as a perturbation. The zeroth order mixing
matrix is
\begin{eqnarray}
U_{l}^F=\left(\begin{array}{ccc}1&0&0\\ 0&c_F&-s_F\\0&
    s_F&c_F \end{array}\right),\label{non_dem_triple_doublet_Ul_T_gg}
\end{eqnarray}
where
\begin{eqnarray}
\tan2\theta_F=\frac{2T_2}{T_3+T_1}=\frac{\sqrt{2}<\chi_2>}{<\chi_3>+<\chi_1>}.
\end{eqnarray}
Note that $\theta_F$ is equal to $\theta_{A,B}$ in the limit
$S\rightarrow 0$ and that $U_l^F$ has the same form as $U_l^B$. When
one includes the perturbation in
(\ref{non_dem_double_triplet_approximate_ml_mldag}), $U_l^F$ will acquire some corrections.
Ultimately, however, the gross structure of $U_l$ in this case
will be set by $U_l^F$. Thus the doublet-triplet case with $T\gg D$ is
very similar to the singlet-triplet case with $T\gg S$. The singlet-triplet case
with $T\gg S$ allows only $2-3$ mixing in the charged lepton
sector. The doublet
provides more free parameters and allows sub-dominant mixing to occur
between the 2-3 states and the first state.

We now consider an example of the neutrino mixing analysis. We shall
now refer to $\Delta_1$, $\Delta_2$ and $\Delta_3$ as the singlet, doublet
and triplet scalars, labelling them by their $SU_H(2)$
transformation properties. The most general mass matrix is
\begin{eqnarray}
M_{\nu}=\left(\begin{array}{ccc} s_1&d_1& d_2
    \\d_1&t_1&t_2\\d_2&
    t_2&t_3 \end{array}\right),\label{non_dem_defining_mnu_form_mu_tau}
\end{eqnarray}
and we investigate the dominant mixing that results from
hierarchies between the elements labelled $s$, $d$ and $t$. As an
example we consider the case $d\gg s$. Retaining the singlet and doublet scalars results in the neutrino mass
matrix
\begin{eqnarray}
M_{\nu}=\left(\begin{array}{ccc} s_1&d_1& d_2
    \\d_1&0&0\\d_2&
    0&0 \end{array}\right).\label{non_dem_singlet_doublet_form_mu_tau}
\end{eqnarray}
This matrix has two non-zero eigenvalues
\begin{eqnarray}
\lambda_\pm=\frac{1}{2}\left(s_1\pm\sqrt{s_1^2+4(d_1^2+d_2^2)}\right),
\end{eqnarray}
and one zero eigenvalue. Upon defining
\begin{eqnarray}
N_\pm=\frac{1}{\sqrt{d_1^2+d_2^2 +\lambda_\pm^2}},
\end{eqnarray}
one may write the mixing matrix as
\begin{eqnarray}
U_{\nu}^b=\left(\begin{array}{ccc} N_+\lambda_+&N_+d_1&N_+d_2\\N_-\lambda_-&N_-d_1&N_-d_2\\0&
    \frac{-d_2}{\sqrt{d_1^2+d_2^2}}&\frac{d_1}{\sqrt{d_1^2+d_2^2}} \end{array}\right).\label{non_dem_triple_doubet_Unu_general}
\end{eqnarray}
This matrix gives
\begin{eqnarray}
U_{\nu}^bM_\nu U_{\nu}^{bT}=\mathrm{diag}(\lambda_+,\lambda_-,0).
\end{eqnarray}
If one neglects $s_1$, ie $d\gg s_1$, the non-zero eigenvalues reduce to
$\lambda_\pm=\pm\sqrt{d_1^2+d_2^2}$. In this limit the mass matrix
(\ref{non_dem_singlet_doublet_form_mu_tau})
displays the family symmetry $L'=L_e-L_\mu-L_\tau$ discussed
earlier. One may define
\begin{eqnarray}
c_c(s_c)=\frac{d_1(d_2)}{\sqrt{d_1^2+d_2^2}},
\end{eqnarray}
and write the mixing matrix as
\begin{eqnarray}
U_{\nu}^c=\left(\begin{array}{ccc}
    \frac{1}{\sqrt{2}}&\frac{c_c}{\sqrt{2}}&\frac{s_c}{\sqrt{2}}\\-\frac{1}{\sqrt{2}}&\frac{c_c}{\sqrt{2}}&\frac{s_c}{\sqrt{2}}\\0&
    -s_c&c_c \end{array}\right).\label{non_dem_triple_doubet_Unu_dgg}
\end{eqnarray}
Observe that
(\ref{non_dem_triple_doubet_Unu_dgg}) displays the exact bi-maximal
form for $d_1=d_2$.
%--------------------------------------------------------------%
\subsection{The PMNS Mixing
  Matrix\label{subsec:non_dem_pmns_matrix_mu_tau}}
We now present the main results of our study, namely the PMNS mixing
matrices $U=U_lU_\nu^\dagger$ emerging from the hierarchies assumed in
the charged lepton and neutrino mass matrices.

The results are presented in the
table below, with the columns of this table containing the following information. The first column
gives the hierarchy assumed in the charged lepton mass matrix. The
second column contains
the leading order rotation of the left-chiral
charged leptons, $U_l$. The
third
column contains the hierarchy assumed in the neutrino mass matrix. The
leading order rotation for the neutrinos, $U_\nu$, is presented in the
fourth column and some important features of the resulting PMNS matrix
are noted in the final column.

The features of a given mixing matrix pointed out in the final column
fall into three categories.\\
$\bullet$ If a given PMNS matrix possesses an entry which is 
zero or near unity in a location which disagrees with the
experimentally observed value this is pointed out to rule out
the matrix.\\
$\bullet$ When a PMNS matrix is not ruled out but makes no specific
predictions we point this out. These matrices are effectively too
general to produce any distinct signature of the horizontal
symmetry.\\
$\bullet$ Features of a given
PMNS matrix which agree well with experiment are also pointed
out. These include $U_{e3}=0$ and
predicted non-zero mixing angles. 

When a rotation appears for the first
time in the table its form is explicitly shown. We use distinct labels for the mixing angles in the
different cases. Many of these angles are not explicitly shown in
the text, however we wish to make it clear that these angles are in general
distinct. The angles not shown correspond to less
interesting
scenarios and should their exact form be required they may readily be
determined. 

As noted earlier, the structure of the charged
lepton rotations $U_{l}^B$ and $U_l^F$ are identical, meaning that the
PMNS matrices which depend on either $U_{l}^B$ or $U_l^F$ have the
same structure. We thus combine these cases together in the table
found below.

\begin{widetext} 
\begin{tabular}{|c|c|c|c|c|}\hline
& & & & \\
$\begin{array}{c}\mathrm{Hierarchy\ In}\\ M_l\end{array}$&
$\begin{array}{c}\mathrm{Charged\ Lepton}\\\mathrm{Mixing}\
  U_l\end{array}$ & $\begin{array}{c}\mathrm{Hierarchy\ In}\\
 M_\nu\end{array}$ & $\begin{array}{c}\mathrm{Neutrino\ Mixing\ }U_\nu^\dagger\end{array}$&$\begin{array}{c}\mathrm{Comments\
on}\\ U=U_lU_\nu^\dagger\end{array}$ \\& & & &\\\hline
%-----------------------------------------------%
 & & & & \\
$S_2\gg T, S$& $U_{l}^A=\left(\begin{array}{ccc} 0&c_A&-s_A\\0&
    s_A&c_A \\1&0&0\end{array}\right)$ &$\begin{array}{c}t\gg s,\\t\gg d\end{array}$ &$U_{\nu}^{a\dagger}=\left(\begin{array}{ccc}
    1&0&0\\0&c_a&-s_a\\0&
    s_a&c_a \end{array}\right)$&$\begin{array}{c}U_{\tau 2}=U_{\tau3}=\\U_{e1}=U_{\mu1}=0\end{array}$  \\ & & & &
\\\hline
%------------------------------------------------%
 & & & & \\
$S_2\gg T, S$& $U_{l}^A$&$\begin{array}{c}s\gg
  t\end{array}$ &$U_{\nu}^{a'\dagger}=\left(\begin{array}{ccc}0&0&1\\c_a&s_a&0\\
    -s_a&c_a&0 \end{array}\right)$&$\begin{array}{c}\mathrm{only\ }\theta_{12}\ne0,\\\theta_{12} \mathrm{\ remains\
    general}\end{array}$  \\ & & & &
\\\hline
%------------------------------------------------%
& & & & \\
$S_2\gg T, S$& $U_{l}^A$& $\begin{array}{c}d\gg s,\\d\gg t\end{array}$&$U_{\nu}^{c\dagger}=\left(\begin{array}{ccc}
    \frac{1}{\sqrt{2}}&-\frac{1}{\sqrt{2}}&0\\\frac{c_c}{\sqrt{2}}&\frac{c_c}{\sqrt{2}}&-s_c\\\frac{s_c}{\sqrt{2}}&\frac{s_c}{\sqrt{2}}&c_c \end{array}\right)$&$\begin{array}{c}U_{\tau3}=0\end{array}$\\ & & & &\\\hline
%------------------------------------------------%
& & & & \\
$S_2\gg T, S$& $U_{l}^A$&$s\gg
d$&$U_{\nu}^{d\dagger}=\left(\begin{array}{ccc}0&-\frac{\sqrt{d_1^2+d_2^2}}{s_1}&1\\
    -s_c&s_c&\frac{d_1}{s_1}\\c_c&s_c&\frac{d_2}{s_1}\end{array}\right)$&$\begin{array}{c}U_{\tau1}=0\end{array}$\\ & & & &\\\hline
%------------------------------------------------%
& & & & \\ 
$\begin{array}{c}T\gg S,S_2,\\ T,S\gg S_2\\\\\mathrm{or}\\\\T\gg D
 \end{array}$&$\begin{array}{c}U_{l}^{B}=\left(\begin{array}{ccc}1&0&0\\ 0&c_B&-s_B\\0&
    s_B&c_B \end{array}\right)\\ \mathrm{or}\\U_l^F=\left(\begin{array}{ccc}1&0&0\\ 0&c_F&-s_F\\0&
    s_F&c_F \end{array}\right)\end{array}$&$\begin{array}{c}t\gg s,\\t\gg d\end{array}$ &$U_{\nu}^{a\dagger}$
&$\begin{array}{c}\mathrm{only}\ 
  \theta_{23}\ne 0,\\\theta_{23} \mathrm{\ remains\
    general}\end{array}$\\ & & & &\\\hline
%------------------------------------------------%
& & & & \\ 
$\begin{array}{c}T\gg S,S_2,\\ T,S\gg S_2\\\mathrm{or}\\T\gg D
  \end{array}$&$\begin{array}{c}\\U_{l}^{B}\\\mathrm{or}\\U_l^F\end{array}$&$\begin{array}{c}s\gg
  t\end{array}$ &$U_{\nu}^{a'\dagger}$
&$\begin{array}{c}U_{e3}=1\end{array}$\\ & & & &\\\hline
%-----------------------------------------------%
& & & & \\ 
$\begin{array}{c}T\gg S,S_2,\\ T,S\gg S_2\\\mathrm{or}\\T\gg D
  \end{array}$&$\begin{array}{c}\\U_{l}^{B}\\\mathrm{or}\\U_l^F\end{array}$&$\begin{array}{c}d\gg s,\\d\gg
  t\end{array}$&$U_{\nu}^{c\dagger}$&$\begin{array}{c}U_{e3}=0\\\tan\theta_{12}=1\end{array}$\\ & & & &\\\hline
%-------------------------------------------------%
& & & & \\ 
$\begin{array}{c}T\gg S,S_2,\\ T,S\gg S_2\\\mathrm{or}\\T\gg D
 \end{array}$&$\begin{array}{c}\\U_{l}^{B}\\\mathrm{or}\\U_l^F\end{array}$&$s\gg
d$&$U_{\nu}^{d\dagger}$&$\begin{array}{c}U_{e1}=0\\U_{e3}=1\end{array}$\\
& & & &\\\hline\end{tabular}
%-------------------------------------------------%
%------------------------------------------------%
\begin{tabular}{|c|c|c|c|c|}\hline
& & & & \\
$\begin{array}{c}\mathrm{Hierarchy\ In}\\ M_l\end{array}$&
$\begin{array}{c}\mathrm{Charged\ Lepton}\\\mathrm{Mixing}\
  U_l\end{array}$ & $\begin{array}{c}\mathrm{Hierarchy\ In}\\
  M_\nu\end{array}$ & $\begin{array}{c}\mathrm{Neutrino\ Mixing\
  }U_\nu^\dagger\end{array}$&$\begin{array}{c}\mathrm{Comments\
on}\\ U=U_lU_\nu^\dagger\end{array}$ \\& & & &\\\hline
%-------------------------------------------------%
& & & & \\$\begin{array}{c}D\gg S\\\mathrm{or}\\D\gg T, \\\mathrm{with\ }\\D_1^2+D_2^2\gg D_3^2+D_4^2\end{array}$&$U_{l}^D=\left(\begin{array}{ccc} 0&-s_D&c_D\\1&
    0&0\\0&c_D&s_D \end{array}\right)$&$\begin{array}{c}t\gg s,\\t\gg
    d\end{array}$
    &$U_{\nu}^{a\dagger}$&$\begin{array}{c}U_{e1}=U_{\mu2}=\\U_{\mu3}=U_{\tau1}=0\end{array}$\\
& & & &\\\hline
%----------------------------------------------%
& & & & \\$\begin{array}{c}D\gg S\\\mathrm{or}\\D\gg T, \\\mathrm{with\ }\\D_1^2+D_2^2\gg D_3^2+D_4^2\end{array}$&$U_{l}^D$&$\begin{array}{c}s\gg
  t\end{array}$
    &$U_{\nu}^{a'\dagger}$&$\begin{array}{c}U_{\mu1}=U_{\mu2}=\\U_{e3}=U_{\tau
    3}=0\end{array}$\\
& & & &\\\hline
%-------------------------------------------------%
& & & & \\$\begin{array}{c}D\gg S\\\mathrm{or}\\D\gg T,
  \\\mathrm{with\ }\\D_1^2+D_2^2\gg
  D_3^2+D_4^2\end{array}$&$U_{l}^D$&$\begin{array}{c}d\gg s,\\d\gg
  t\end{array}$&$U_{\nu}^{c\dagger}$&$U_{\mu3}=0$\\
& & & &\\\hline
%-------------------------------------------------%
& & & & \\$\begin{array}{c}D\gg S\\\mathrm{or}\\D\gg T,
  \\\mathrm{with\ }\\D_1^2+D_2^2\gg
  D_3^2+D_4^2\end{array}$&$U_{l}^D$&$s\gg
d$&$U_{\nu}^{d\dagger}$&$U_{\mu1}=0$\\
& & & &\\\hline
%-----------------------------------------------%
& & & & \\$\begin{array}{c}D\gg S\\\mathrm{or}\\D\gg T,
  \\\mathrm{with\ }\\D_3^2+D_4^2\gg
  D_1^2+D_2^2\end{array}$&$U_{l}^E=\left(\begin{array}{ccc} 0&-s_E&c_E\\0&
    c_E&s_E\\1&0&0 \end{array}\right)$&$\begin{array}{c}t\gg s,\\t\gg
  d\end{array}$
  &$U_{\nu}^{a\dagger}$&$\begin{array}{c}U_{e1}=U_{\mu1}=\\U_{\tau2}=U_{\tau3}=0\end{array}$\\
& & & &\\\hline
%-------------------------------------------------%
& & & & \\$\begin{array}{c}D\gg S\\\mathrm{or}\\D\gg T,
  \\\mathrm{with\ }\\D_3^2+D_4^2\gg
  D_1^2+D_2^2\end{array}$&$U_{l}^E$&$\begin{array}{c}s\gg
  t\end{array}$
  &$U_{\nu}^{a'\dagger}$&$\begin{array}{c}\mathrm{only\ }\theta_{12}\ne0,\\\theta_{12} \mathrm{\ remains\
    general}\end{array}$\\
& & & &\\\hline
%-------------------------------------------------%
& & & & \\$\begin{array}{c}D\gg S\\\mathrm{or}\\D\gg T,
  \\\mathrm{with\ }\\D_3^2+D_4^2\gg
  D_1^2+D_2^2\end{array}$&$U_{l}^E$&$\begin{array}{c}d\gg s,\\d\gg
  t\end{array}$&$U_{\nu}^{c\dagger}$& $U_{\tau3}=0$\\
& & & &\\\hline
%-------------------------------------------------%
& & & & \\$\begin{array}{c}D\gg S\\\mathrm{or}\\D\gg T,
  \\\mathrm{with\ }\\D_3^2+D_4^2\gg
  D_1^2+D_2^2\end{array}$&$U_{l}^E$&$s\gg
d$&$U_{\nu}^{d\dagger}$& $U_{\tau1}=0$\\
& & & &\\\hline
%-------------------------------------------------%
\end{tabular}\\

\end{widetext}
To discuss the PMNS matrices constructed in the above table we define
$U^{Xy}=U^X_lU_\nu^{y\dagger}$. Of the sixteen matrices
in the table one may discount twelve due to the presence
of leading order zero entries in undesirable locations. Two of the
remaining PMNS matrices, $U^{Aa'}$ and $U^{Ea'}$,  predict only $\theta_{12}\ne 0$ and offer no
insight into what the value of $\theta_{12}$ should be. Evidently, it is not
desirable to consider the near
maximal value of $\theta_{23}$ observed experimentally as the result
of a perturbative correction to these matrices. 

The matrices $U^{Fa}$ and $U^{Ba}$
contain $\theta_{23}$ as the only non-zero mixing
angle. It is possible to construct realistic PMNS matrices as
perturbations upon their existing structures. However, these matrices
make no prediction as to what value $\theta_{23}$ should take
and in the absence of an explicit model that generates VEV hierarchies
amongst the VEV's $t_{123}$ or $T_{123}$ they have little predictive
power.

The matrices $U^{Fc}$ and $U^{Bc}$ prove to be the most
interesting. They possess the
same leading order structure, each containing a zero in the $e3$
position. The neutrino mixing matrix
$U_\nu^c$ results from a mass matrix of the form
\begin{eqnarray}
M_{\nu}=\left(\begin{array}{ccc} 0&d_1& d_2
    \\d_1&0&0\\d_2&
    0&0 \end{array}\right).\label{non_dem_doublet_form_mu_tau_discussion}
\end{eqnarray}
This displays the $L'=L_e-L_\mu-L_\tau$ symmetry, leading to an
inverted hierarchy with $\Delta m_{12}^2=0$. Perturbations are
required to produce a splitting of order $\Delta
m_{12}^2\sim10^{-5}$~eV$^2$ and a non-maximal solar mixing
angle. These perturbations may come
from the charged lepton mass matrix or the neutrino mass matrix.

Consider first the charged lepton sector. The matrices $U^{Fc}$ and $U^{Bc}$ occur when the triplet scalar
entries dominate the charged lepton mass matrix,
\begin{eqnarray}
M_{l}=\left(\begin{array}{ccc} 0&0& 0
    \\0&T_2&T_1\\0&
    T_3&-T_2 \end{array}\right).\label{non_dem_triplet_only_discussion_mu_tau}
\end{eqnarray}
If the
singlet entries are included one obtains
\begin{eqnarray}
M_{l}=\left(\begin{array}{ccc} S_2&0& 0
    \\0&S+T_2&T_1\\0&
    T_3&S-T_2 \end{array}\right),\label{non_dem_triplet_singlet_discussion_mu_tau}
\end{eqnarray}
so that $m_e=S_2$. This does not alter the solar mixing angle.
If one instead includes the doublet
entries in $M_l$, 
\begin{eqnarray}
M_{l}=\left(\begin{array}{ccc} 0&D_3& D_4
    \\D_1&T_2&T_1\\D_2&
    T_3&-T_2 \end{array}\right),\label{non_dem_triplet_doublet_discussion_mu_tau}
\end{eqnarray}
one obtains $m_e\sim D$. A correction
of order \(1^\circ\) to the solar mixing angle $\theta_{12}$ results, which is much too small to
reach the experimentally favoured value of $\theta_{12}\sim
34^\circ$~\cite{Choubey:2005rq}.

One can consider the addition of both the
singlet and doublet entries to the charged lepton mass matrix. However
the electron
mass develops a dependence on both $D$ and $S$ and the mixing
angles emerging from the charged lepton sector remain too small under
our assumed hierarchies. 

If $d_1\sim d_2$ the neutrino mixing matrix has
the bi-maximal form. A previous work has studied the forms of the
charged lepton mixing matrix which are compatible with the
experimentally observed PMNS matrix values when one assumes bi-maximal neutrino
mixing~\cite{Frampton:2004ud}. The failure of the charged lepton mass
matrices to
adequately perturb the bi-maximal form of the neutrino mixing matrix in
the present analysis
reflects the disagreement between the forms of $M_l$ encompassed by
the present study and those
obtained in~\cite{Frampton:2004ud}. 

We now consider the addition of perturbations  to the neutrino
mass matrix. One can readily show that the addition of only $s_1$ to
$M_\nu$ does not permit realistic values for $\Delta m_{12}^2$
and $\theta_{12}$ simultaneously. We shall consider the case where the singlet and triplet entries are included in the neutrino mass matrix. We are interested in the minimum additional number of scalar multiplets required to produce a realistic neutrino spectrum. However the case where only the triplet entries are included will readily be revealed in the appropriate limit of this more general case. We shall work in the basis where the charged lepton mass matrix has been diagonalized by the matrix $U_l^F$. We parameterise the neutrino mass matrix as 
\begin{eqnarray}
M_{\nu}&=&m_0\left(\begin{array}{ccc} 0 &c& -s
    \\c&0&0\\-s&
    0&0 \end{array}\right)+m_0\left(\begin{array}{ccc} -w &0& 0
    \\0&-x&\frac{y}{\sqrt{2}}\\0&
    \frac{y}{\sqrt{2}}&z \end{array}\right)\nonumber\\
&\equiv&m_0\left\{M_0 + M_1\right\},\label{non_dem_triplet_doublet_form_mu_tau_defining_perturbation}
\end{eqnarray}
where $c=\cos\theta_{23}$ and $s=\sin\theta_{23}$ will turn out to be
the cosine and sine of the atmospheric mixing angle. The mass scale of
the eigenvalues is set by $m_0\sim d$ and the entries in $M_1$
labelled $x$, $y$ and $z$ are of order $t/m_0$ whilst $w=s_1/m_0$. To
first order in the small parameters $x,y,z$ and $w$ the ratio of mass
squared differences is
\begin{eqnarray}
\frac{\Delta m_{12}^2}{\Delta m_{23}^2}\approx\frac{2(2\kappa-w)}{1-(2\kappa-w)},\label{non_dem_perturbed_mass_2_ratio}
\end{eqnarray}
where we have introduced
\begin{eqnarray}
\kappa&=&\frac{1}{2}(s^2 z-c^2 x) -\frac{y}{2\sqrt{2}}\sin2\theta_{23},\nonumber\\
\end{eqnarray}
and it will prove useful to define
\begin{eqnarray}
\xi&=& 
\frac{1}{2\sqrt{2}}(x+z)\sin2\theta_{23}-\frac{y}{2}\cos2\theta_{23}.
\end{eqnarray}
The first order mixing matrix takes the form
\begin{widetext}
\begin{eqnarray}
U=\left(\begin{array}{ccc}\frac{1}{\sqrt{2}} +f & \frac{1}{\sqrt{2}} -f & \sqrt{2}\xi\\ -\frac{c}{\sqrt{2}} +cf-s\kappa&\frac{c}{\sqrt{2}} +cf-s\kappa& s \\ \frac{s}{\sqrt{2}} -sf-c\kappa & -\frac{s}{\sqrt{2}} -sf-c\kappa& c \end{array}\right),
\end{eqnarray}
\end{widetext}
where
\begin{eqnarray}
f&=&\frac{1}{4\sqrt{2}}(2\kappa+w).
\end{eqnarray}
The neutrino mixing angles are
\begin{eqnarray}
\tan\theta_{13}&\approx&\sqrt{2}\xi\label{non_dem_reactor_angle_triplet_singlet_perturbations},\\
\tan\theta_{23}&\approx&\frac{s}{c}\label{non_dem_atm_angle_triplet_singlet_perturbations},\\
\tan\theta_{12}&\approx& 1-\frac{1}{2}(2\kappa+w)\label{non_dem_solar_angle_triplet_singlet_perturbations}.
\end{eqnarray}
Consider first the case where no singlet entry is present in the
neutrino mass matrix, namely the limit $w\rightarrow 0$ in
$M_1$. One then has
\begin{eqnarray}
\frac{\Delta m_{12}^2}{\Delta m_{23}^2}\approx\frac{2\kappa}{1-2\kappa},\label{non_dem_perturbed_singlet_only_mass_2_ratio}
\end{eqnarray}
and
\begin{eqnarray}
\tan\theta_{12}\approx 1+\kappa.\label{non_dem_theta_12_no_singlet}
\end{eqnarray}
The current allowed $3\sigma$ ranges for the mass squared differences are~\cite{Valle:2005ai}
\begin{eqnarray}
7.1\le\Delta m_{12}^2/(10^{-5}eV^2)\le 8.9,\nonumber\\
1.4\le\Delta m_{23}^2/(10^{-3}eV^2)\le3.3,\label{rev_non_valle_mass_2_3_sigma_bounds}
\end{eqnarray}
and the mixing angles lie in the ranges
\begin{eqnarray}
0.24\le\sin^2\theta_{12}\le0.4,\nonumber\\
0.34\le\sin^2\theta_{23}\le0.68,\nonumber\\
\sin^2\theta_{13}\le 0.047.
\end{eqnarray}
Equation (\ref{non_dem_perturbed_singlet_only_mass_2_ratio}) requires the single parameter $\kappa$ to be small (of order $10^{-2}$), thereby rendering the required deviation from maximal solar mixing unrealisable in concordance with equation (\ref{non_dem_theta_12_no_singlet}).

When both the singlet and triplet entries are included as
perturbations realistic values for both the ratio of mass squared
differences (\ref{non_dem_perturbed_mass_2_ratio}) and the solar
mixing angle (\ref{non_dem_solar_angle_triplet_singlet_perturbations})
can be obtained. This results from the fact that the solar mixing
angle depends on the linear combination of parameters $(2\kappa+w)$ in
(\ref{non_dem_solar_angle_triplet_singlet_perturbations}) whilst the
ratio of mass squared differences in
(\ref{non_dem_perturbed_mass_2_ratio}) depends on the linearly
independent combination $(2\kappa-w)$. Assuming a maximal value for
$\theta_{23}$ the experimental range of $\theta_{12}$ values are
obtained for $w$ and $\kappa$ in the approximate ranges
$|w|\approx[0.18, 0.44]$ and $|\kappa| \approx[0.09,0.22]$.

This result is similar to that obtained in the work by Leontaris,
Rizos and Psallidas in reference~\cite{non_dem_lprime_symmetry} (and
more recently~\cite{Leontaris:2005gm}). In that work lepton mass
matrices with leading order structures identical to
(\ref{non_dem_doublet_form_mu_tau_discussion}) and
(\ref{non_dem_triplet_singlet_discussion_mu_tau}) were obtained by
augmenting the standard model with an anomalous $U(1)$ symmetry.

As the spectrum is inverted, with $m_3= 0$~eV to zeroth order in the small parameters, one has
\begin{eqnarray}
m_0\simeq \sqrt{\Delta m_{23}^2},\nonumber
\end{eqnarray}
giving the $3\sigma$ range
\begin{eqnarray}
m_0/(10^{-2}\mathrm{eV})\in[3.7,5.7].\label{non_dem_neutrino_mass_scale}
\end{eqnarray}
The effective mass in neutrinoless double beta decay may be expressed as
\begin{eqnarray}
|<m>|&=&|\sum_{i} U_{ei}^2 m_{i}|\nonumber\\
&=&m_0 w,
\end{eqnarray}
where $i=1,2,3,$ labels the neutrino mass eigenstates. Using the allowed range for $w$ and (\ref{non_dem_neutrino_mass_scale}) one obtains
\begin{eqnarray}
<m>/(10^{-2}\mathrm{eV})\in[0.7,2.5],\label{non_dem_m_ee_range}
\end{eqnarray}
which puts the effective neutrinoless double beta decay mass just
below current experimental bounds (see eg~\cite{Valle:2005ai}). Thus
if the present framework is valid one is likely to observe
neutrinoless double beta decay in the next generation of experiments,
which will reach an accuracy of  $<m>\sim\mathrm{few}$~$\times
10^{-2}$~eV~\cite{Bilenky:2002aw,Bilenky:2005bq}. The mixing matrix
element $U_{e3}=\sqrt{2}\xi$ is not uniquely determined in the present
framework. Taking the atmospheric mixing angle to be maximal gives
$U_{e3}= (x+z)/2$, with the small parameters $x$ and $z$ not
sufficiently constrained to allow a unique prediction.
%--------------------------------------------------------------%
\subsection{Summary\label{subsec:non_dem_summary_mu_tau}}
We have found that a realistic lepton spectrum and PMNS
matrix are obtained when couplings to the scalars $\chi$ and
$\Delta_2$ dominate the charged lepton and neutrino mass matrices
respectively. The required fine structure necessitates the inclusion
of mass matrix entries resulting from couplings to $\Delta_1$ and
$\Delta_3$ in the neutrino sector and either $\phi$ or $\Phi$ in the
charged lepton sector. Thus a minimum of five (out of the possible
six) scalar multiplets are necessary to reproduce the experimentally
observed lepton sector within the present framework. The neutrinos are
required to display an inverted hierarchy, with the leading order
neutrino mass matrix displaying the non-standard lepton number
symmetry $L'=L_e-L_\mu-L_\tau$. Should the neutrino mass spectrum turn
out to be inverted the horizontal symmetry $SU_H(2)$ may
explain why nature displays the approximate symmetry $L'$.
%---------------------------------------------------------%
\section{Scalars And Horizontal Gauge Bosons\label{sec:non_dem_scalars_bosons}}
We wish to make a few remarks regarding the scalar content required
for a realistic model and the mass scale of the horizontal gauge
bosons. Though it is beyond the scope of the present work to construct
and analyse a specific scalar potential, a comment on the scale of the
VEV's obtained by the $SU_L(2)$ triplet fields is in order. 

It is known that $SU_L(2)$ triplet fields may acquire naturally light
VEV's if the scalar potential contains a term linear in the given
field~\cite{Ma:1998dx}. In the case of the SM augmented to include one
$SU_L(2)$ triplet field $\Delta$, the field $\Delta$ acquires a VEV of
order $\mu v^2/M_\Delta^2$, where $v$ denotes the VEV of the SM Higgs
field $\Phi'$, $M_\Delta$ is the mass of the field $\Delta$ and $\mu$
denotes the coefficient of the scalar potential term linear in
$\Delta$, ie $\mu\Delta \Phi' \Phi'\subset V$, where $V$ is the scalar
potential. Thus provided $\Delta$ is heavy enough its VEV will be
suppressed relative to the electroweak scale.

Within the framework employed in the present study we have seen that a
realistic neutrino sector requires three $SU_L(2)$ triplet fields,
$\Delta_1$, $\Delta_2$ and $\Delta_3$, which form singlet, fundamental
and triplet representations of $SU_H(2)$ respectively. Evidently we
would like to ensure that the potential contains terms linear in these
fields to allow them to develop VEV's below the electroweak scale. The
minimal scalar content required to produce a realistic charged lepton
spectrum consists of the $SU_L(2)$ doublet fields $\phi$ and $\chi$,
which form respectively singlet and triplet representations of
$SU_H(2)$. With these five fields the potential will contain the
terms:
\begin{eqnarray}
V\supset \Delta_{1}(\mu_{11} \phi \phi +\mu_{13} \chi \chi)+\Delta_{3}(\mu_{31}\phi \chi +\mu_{33} \chi \chi),
\end{eqnarray}
where the $\mu$'s are all dimension-full coupling constants. The above
terms, being linear in $\Delta_1$ and $\Delta_3$, should allow these
fields to naturally develop sub electroweak scale VEV's. Without the
inclusion of the field $\Phi$, which transforms as $(2,2)$ under
$SU_L(2)\otimes SU_H(2)$, the potential does not contain a term linear
in $\Delta_2$. The addition of $\Phi$ to the scalar spectrum results
in additional potential terms, including
\begin{eqnarray}
\Delta_{2}\mu_{22}\Phi \chi,
\end{eqnarray}
which may allow $\Delta_2$ to obtain a VEV suppressed relative to the
electroweak scale. Although a realistic charged lepton spectrum does
not require the field $\Phi$, this field may be necessary to ensure the
potential contains an appropriate term linear in $\Delta_2$. 

In the work of KM the $SU_H(2)$ symmetry
breaking scale is related to the light neutrino masses via the see-saw
mechanism, namely $m_\nu\sim v^2/M_H$, where $M_H$ denotes the scale
of horizontal symmetry breaking and $v$ is the electroweak
scale. Assuming order one Yukawa couplings constrains $M_H$ to be
roughly of order $10^{14}$~GeV. Consequently the horizontal gauge
bosons become unobservably heavy.

In the present work, if an $SU_H(2)$ doublet of right-chiral neutrinos
is included to remove the global anomaly, the see-saw contribution to
the mass of the light neutrinos is required to be sub-dominant. Thus
the lower bound on the scale of horizontal symmetry breaking becomes
even more severe. Note that one only requires $SU_H(2)$ to be broken
to $U(1)$ to generate heavy Majorana masses and we have no $a$
$priori$ reason to break the $U(1)$ subgroup at the scale
$M_H$. However it may prove difficult to retain only mild hierarchies
amongst the VEV's of a given scalar multiplet with a non-zero
horizontal charge (e.g. amongst $<\chi_1>$, $<\chi_2>$ and $<\chi_3>$)
if this symmetry is broken at a vastly lower scale (see for example
the discussion in~\cite{Kuchimanchi:2002yu}). It is beyond the scope
of the present work to pursue this matter further.

The constraints on the horizontal symmetry breaking scale are much
less severe if the horizontal symmetry communicates with a mirror
sector. In this case the model need not include right chiral neutrinos
and no see-saw induced connection between the mass of the light
neutrinos and the scale of horizontal symmetry breaking arises.

The horizontal gauge bosons do not couple directly to quarks so that many of the usual bounds on additional neutral gauge bosons do not apply. The horizontal gauge bosons couple to the charged leptons as follows:
\begin{eqnarray}
-\mathcal{L}&=&g_H Z_\mu' \left\{\bar{L}_L\gamma^\mu U_1 L_L + \bar{L}_R\gamma^\mu V_1L_R\right\}+\nonumber\\
& &\frac{g_H}{\sqrt{2}}Z_\mu'\left\{\bar{L}_L\gamma^\mu U_2 L_L+ \bar{L}_R\gamma^\mu V_2 L_R\right\} +\mathrm{h.c.},\nonumber  
\end{eqnarray} 
where we use the label $Z'$ generically to denote a horizontal gauge boson and $g_H$ is the horizontal coupling constant. The horizontal charged lepton mixing matrices take the form
\begin{eqnarray}
U_1&=&U_lC_1U_l^\dagger,\nonumber\\
U_2&=&U_lC_2U_l^\dagger,\nonumber\\
V_1&=&V_l C_1V_l^\dagger,\nonumber\\
V_2&=&V_lC_2V_l^\dagger,
\end{eqnarray}
where $U_l$ and $V_l$ are the left- and right-chiral charged lepton mixing matrices respectively,
\begin{eqnarray}
U_lM_lM_l^\dagger U_l^\dagger=\mathrm{diag}(m_e^2,m_\mu^2,m_\tau^2),\nonumber\\
V^\dagger_lM_l^\dagger M_lV_l=\mathrm{diag}(m_e^2,m_\mu^2,m_\tau^2),
\end{eqnarray}
and
\begin{eqnarray}
C_1=\mathrm{diag}(0,1,-1),\nonumber\\
C_2=\left(\begin{array}{ccc}0&0&0\\0&0&1\\0&0&0\end{array}\right).
\end{eqnarray}
The matrices $C_{1,2}$ reflect the fact that there is a basis in which
only two generations of leptons couple to the horizontal bosons and
are proportional to linear combinations of the Pauli matrices.

When the triplet entries dominate the charged lepton mass matrix one has $U_l=U_l^F$ and $V_l=P' U_l^{F\dagger}$ to leading order, with
\begin{eqnarray}
P'&=&\left(\begin{array}{ccc}1&0&0\\0&0&1\\0&-1&0\end{array}\right).
\end{eqnarray}
In the absence of the field $\Phi$ the horizontal gauge bosons only
couple to the muon and tauon generations of leptons. If the field
$\Phi$ is included charged lepton mixing will couple the electron
generation to the horizontal gauge bosons. Thus in principle the
horizontal gauge bosons could be observed in $e^+ e^-$
collisions. However the hierarchies assumed in the present study mean
that the associated mixing angle will be of order $(m_e/m_\mu)^2$ or
smaller. To see this, note that under the assumed hierarchies the
horizontal mixing matrices will take the symbolic forms
\begin{eqnarray}
U_1,V_1&\sim& \left(\begin{array}{ccc}\epsilon^2 &\epsilon \delta&\epsilon \delta \\\epsilon \delta&1&\epsilon \delta\\\epsilon \delta&\epsilon\delta&-1\end{array}\right),
\end{eqnarray}
and
\begin{eqnarray}
U_2,V_2&\sim& \left(\begin{array}{ccc}\epsilon^2 &\epsilon \delta&\epsilon \delta \\\epsilon \delta&\delta^2&\delta^2\\\epsilon \delta&\delta^2&\delta^2\end{array}\right),
\end{eqnarray}
where $\epsilon$ denotes (different) elements of order $m_e/m_\mu$ or smaller, $\delta\sim c_F, s_F$ and $\delta^2\sim c_F^2,s_F^2,c_Fs_F$. Thus the coupling of $e^+e^-$ pairs to the horizontal gauge bosons are suppressed by a factor of $(m_e/m_\mu)^2\sim 10^{-5}$. The flavour changing vertices coupling electrons and muons are suppressed by the larger factor of ($m_e/m_\mu$). One can use the bound on the lepton number violating decay
\begin{eqnarray}
\tau\rightarrow2\mu+ e,
\end{eqnarray}
to constrain the coupling and mass of the horizontal bosons. If we follow~\cite{Shaw:1992gk} and compare the branching ratio for this process to the SM decay
\begin{eqnarray}
\tau\rightarrow \mu +2\nu,
\end{eqnarray}
we find~\cite{shaw_error}
\begin{eqnarray}
\frac{g_H^4}{M_{Z'}^4}\le \frac{B(\tau\rightarrow2\mu e)}{B(\tau\rightarrow\mu2\nu)}\times \left(\frac{m_\mu}{m_e}\right)^2\times \frac{g_W^4}{M_W^4},
\end{eqnarray}
where $g_W$ is the weak coupling constant, $M_{Z'}$ ($M_W$) is the horizontal ($W$) boson mass and we have taken $\delta\sim1$. Using the current bound $B(\tau\rightarrow2\mu e)<1.5\times 10^{-5}$ and $B(\tau\rightarrow\mu2\nu)=0.1736\pm0.0006$~\cite{Eidelman:2004wy} one obtains the weak bound
\begin{eqnarray}
\frac{g_H^4}{M_{Z'}^4}<3.7\times \frac{g_W^4}{M_W^4}.
\end{eqnarray} 
This demonstrates the difficulty in obtaining sensitive bounds on the horizontal bosons from processes involving electrons. The horizontal gauge bosons will give rise to additional tauon decay modes of the type $\tau\rightarrow \mu + invisible$, namely
\begin{eqnarray}
\tau\rightarrow \mu + Z'\rightarrow \mu + \nu_\mu +\nu_\tau,\nonumber\\
\tau\rightarrow \mu + Z'\rightarrow \mu + \nu_\mu' +\nu_\tau'.\label{non_dem_new_tau_decay}
\end{eqnarray}
Here $\nu'$ denotes mirror neutrinos, recalling that the horizontal gauge bosons couple to the mirror sector~\cite{zprime_morror_note}. One may obtain bounds on the horizontal bosons by demanding that
\begin{widetext}
\begin{eqnarray}
\Gamma_{Extra}(\tau\rightarrow\mu+inv)&\le&\Gamma_{Exp}(\tau\rightarrow\mu+inv)-\Gamma_{SM}(\tau\rightarrow\mu+inv),
\end{eqnarray}
\end{widetext}
where $\Gamma_{Extra}(\tau\rightarrow\mu+inv)$ is the width for the new decays (\ref{non_dem_new_tau_decay}), $\Gamma_{Exp}(\tau\rightarrow\mu+inv)$ is the experimentally measured width and $\Gamma_{SM}(\tau\rightarrow\mu+inv)$ is the SM value. Using the measured values~\cite{Eidelman:2004wy},
\begin{eqnarray}
& &B(\tau\rightarrow\mu+inv)=0.1736,\nonumber\\
& &\tau=290.6\times 10^{-15}s,
\end{eqnarray}
gives
\begin{eqnarray}
\Gamma_{Exp}(\tau\rightarrow\mu+inv)=3.93\times10^{-10}~\mathrm{MeV}.
\end{eqnarray}
The SM value may be calculated as~\cite{Tsai:1971vv}
\begin{eqnarray}
\Gamma_{SM}(\tau\rightarrow\mu+inv)=\frac{G_F^2m_\tau^5}{192 \pi^3}F\left(\frac{m_\mu^2}{m_\tau^2}\right)\times r_{EW},
\end{eqnarray}
where $G_F$ is the Fermi constant, $r_{EW}=0.9960$ accounts for the electroweak propagator and radiative corrections~\cite{Marciano:1988vm} and
\begin{eqnarray}
F(x)=1-8x+8x^3-x^4-12x\ln x.
\end{eqnarray}
This gives
\begin{eqnarray}
\Gamma_{SM}(\tau\rightarrow\mu+inv)=3.90\times10^{-10}~\mathrm{MeV},
\end{eqnarray}
leading to the upper limit
\begin{eqnarray}
\Gamma_{Extra}(\tau\rightarrow\mu+inv)\le 3.0\times 10^{-12}~\mathrm{MeV}.
\end{eqnarray}
This translates into the more restrictive bound
\begin{eqnarray}
\frac{g_H^2}{M_{Z'}^2}\le6\times10^{-2}\frac{g_W^2}{M_{W}^2},
\end{eqnarray}
which gives $M_{Z'}\ge320$~GeV for $g_H=g_W$. This bound is low enough
to allow the possible observation of horizontal gauge boson
contributions to processes like $e^+e^-\rightarrow \mu^+\mu^-$ at the
Next Linear Collider~\cite{next_linear_collider}. As the coupling of
the horizontal bosons to $\mu^\pm$ is larger than the couplings to
$e^\pm$, the horizontal bosons would create interesting phenomenology
at a muon collider. Note also that the matrix $V_l$ appears in the
couplings between $Z'$ and the charged leptons. This allows one to
experimentally
reconstruct the matrix $V_l$.  
%-------------------------------------------------------------%
\section{Conclusion\label{sec:non_dem_conclusion}}
We have performed a systematic analysis of the PMNS matrices which
arise when the three generations of leptons are assigned to the
$2\oplus 1$ representation of the horizontal symmetry $SU_H(2)$, and
the neutrino mass matrix results from leptonic couplings to $SU_L(2)$
triplet scalar fields. It was assumed that hierarchies existed amongst
lepton mass matrix elements which result from couplings to scalar
fields with different charges under $SU_H(2)$. Of the sixteen
candidate PMNS matrices which arose in our study it was found that
only one was both predictive and possessed a leading order structure
compatible with experimental observations. The relevant neutrino mass
matrix displayed the symmetry $L_e-L_\mu-L_\tau$ to leading order and
emerged when the contribution to the charged lepton mass matrix by a
scalar forming a $(2,1,3)$ representation of $SU_L(2)\otimes
U_Y(1)\otimes SU_H(2)$ dominated and the $(3,2,2)$ contributions to
the neutrino mass matrix dominated. This PMNS matrix predicted maximal
solar mixing, $U_{e3}=0$ and left the atmospheric mixing angle
unconstrained. It also required the neutrinos to display an inverted
hierarchy. Perturbations to this leading order structure resulted from
contributions to the neutrino mass matrix from additional scalar
multiplets. Experimental data directly constrained the parameters
which entered into the effective mass in neutrinoless double beta
decay. This resulted in the prediction
$<m>/(10^{-2}\mathrm{eV})\in[0.7,2.5]$, which is just below current
experimental bounds.

We also noted that the contribution of $SU_L(2)$ triplet scalar
multiplets to the neutrino mass matrix can only dominate if the
standard see-saw contribution is sub-dominant or not present. In the
first case an $SU_H(2)$ doublet of right-chiral neutrinos may be
included to remove the global anomaly as in the original work of
KM. The sub-dominant nature of the see-saw contribution requires
$SU_H(2)$ to be broken at an unobservably large scale. If there are no
right-chiral neutrinos the global anomaly may be removed by allowing
the horizontal symmetry to communicate with a mirror sector. In this
case the scale of horizontal symmetry breaking may be much lower with
hope of observing resonance behaviour of the horizontal gauge bosons
at $e^+e^-$ colliders operating at TeV energies.
%------------------------------------------------------------------%
\section*{Acknowledgements}
K.M. thanks Raymond Volkas for suggestions regarding an early version of this manuscript and Robert Foot for useful discussions. This work was
supported in part by the Australian Research Council.
%---------------------------------------------------------------------%

%-----------------------------------------------------------------------------%

\begin{thebibliography}{999}
\bibitem{zprime_morror_note}The decay of $Z'$ into mirror neutrinos need
not occur if the mirror symmetry is spontaneously broken. To ensure
that oscillations of the the ordinary neutrinos into mirror neutrinos
are at most sub-dominant on solar system length scales, one requires
either very small mass squared differences between the mirror parity
eigenstates, or a spontaneously broken mirror symmetry to generate
small ordinary-mirror mixing angles.
\end{thebibliography}
\end{document}